\documentclass[doublecol]{epl2} 

\usepackage{subfigure}

\title{On the dark nature of exciton Bose-Einstein condensate}

\author{Monique Combescot\inst{1}\thanks{E-mail: \email{Monique.Combescot@insp.jussieu.fr}} \and Michael N. Leuenberger\inst{2}\thanks{E-mail: \email{mleuenbe@mail.ucf.edu}}}

\institute{                    
  \inst{1} Institut des NanoSciences de Paris, CNRS, Universite Pierre et Marie Curie  - 140 rue de Lourmel, 75015 Paris\\
  \inst{2} NanoScience Technology Center and Dept. of Physics, University of Central Florida - 12424 Research Parkway Suite 400, Orlando, FL 32826
}

\abstract{
We show that for the very same reason that excitons are bright, i.e. coupled to photons, they have a higher energy than dark excitons, even for electrons spatially separated from holes, such as in a double quantum well. Indeed, the same channel which produces the finite electron-hole effective overlap responsible for the absorption and emission of photon allows for Coulomb interband exchange processes, which are nothing but a sequence of virtual recombination and creation of one electron-hole pair. Consequently, this additional repulsive electron-hole Coulomb exchange interaction exists for bright excitons, but not for dark excitons. If we now remember that dark excitons with spins $\pm 2$ are formed in a natural way through carrier exchange between two opposite spin bright excitons, we are led to predict that in a double quantum well sample with one trap -- a configuration appropriate to get high density -- exciton Bose-Einstein condensation should appear when cooling down the sample as a dark spot made of $(\pm 2)$ excitons at the center of the trap.  
}

\begin{document}

\newcommand{\1}{{\bf \scriptstyle 1}\!\!{1}}
\newcommand{\I}{{\rm i}}
\newcommand{\p}{\partial}
\newcommand{\D}{^{\dagger}}
\newcommand{\bx}{{\bf x}}
\newcommand{\br}{{\bf r}}
\newcommand{\bk}{{\bf k}}
\newcommand{\bv}{{\bf v}}
\newcommand{\bp}{{\bf p}}
\newcommand{\bq}{{\bf q}}
\newcommand{\bs}{{\bf s}}
\newcommand{\bu}{{\bf u}}
\newcommand{\bA}{{\bf A}}
\newcommand{\bB}{{\bf B}}
\newcommand{\bE}{{\bf E}}
\newcommand{\bF}{{\bf F}}
\newcommand{\bG}{{\bf G}}
\newcommand{\bI}{{\bf I}}
\newcommand{\bJ}{{\bf J}}
\newcommand{\bK}{{\bf K}}
\newcommand{\bL}{{\bf L}}
\newcommand{\bP}{{\bf P}}
\newcommand{\bQ}{{\bf Q}}
\newcommand{\bS}{{\bf S}}
\newcommand{\bH}{{\bf H}}
\newcommand{\balpha}{\mbox{\boldmath $\alpha$}}
\newcommand{\bsigma}{\mbox{\boldmath $\sigma$}}
\newcommand{\bSigma}{\mbox{\boldmath $\Sigma$}}
\newcommand{\bOmega}{\mbox{\boldmath $\Omega$}}
\newcommand{\bpi}{\mbox{\boldmath $\pi$}}
\newcommand{\bphi}{\mbox{\boldmath $\phi$}}
\newcommand{\bnabla}{\mbox{\boldmath $\nabla$}}
\newcommand{\bmu}{\mbox{\boldmath $\mu$}}
\newcommand{\bepsilon}{\mbox{\boldmath $\epsilon$}}

\newcommand{\iLambda}{{\it \Lambda}}
\newcommand{\cA}{{\cal A}}
\newcommand{\cD}{{\cal D}}
\newcommand{\cF}{{\cal F}}
\newcommand{\cL}{{\cal L}}
\newcommand{\cH}{{\cal H}}
\newcommand{\cI}{{\cal I}}
\newcommand{\cM}{{\cal M}}
\newcommand{\cO}{{\cal O}}
\newcommand{\cR}{{\cal R}}
\newcommand{\cU}{{\cal U}}
\newcommand{\cT}{{\cal T}}
\newcommand{\cV}{{\cal V}}

\newcommand{\be}{\begin{equation}}
\newcommand{\ee}{\end{equation}}
\newcommand{\bea}{\begin{eqnarray}}
\newcommand{\eea}{\end{eqnarray}}
\newcommand{\beqa}{\begin{eqnarray*}}
\newcommand{\eeqa}{\end{eqnarray*}}
\newcommand{\nn}{\nonumber}
\newcommand{\DD}{\displaystyle}

\newcommand{\ba}{\left[\begin{array}{c}}
\newcommand{\baa}{\left[\begin{array}{cc}}
\newcommand{\baaa}{\left[\begin{array}{ccc}}
\newcommand{\baaaa}{\left[\begin{array}{cccc}}
\newcommand{\ea}{\end{array}\right]}

\maketitle

Bose-Einstein condensation of boson-like particles is a fascinating quantum effect \cite{Pitaevskii}. While easy to follow from textbook arguments, it is far harder to physically grasp that  bosonic atoms with zero momentum are the ones which really condensed, since they are not so much different in energy from the ones with small non-zero momentum for samples of macroscopic size. With respect to this point, it has been noted by Nozieres \cite{Nozieres} and also by Leggett \cite{Leggett} that for elementary bosons, the non-fragmentation of Bose-Einstein condensate is greatly helped by interactions, always present in real experiments. Indeed, the possible exchange interactions between two indistinguishable sets of elementary bosons with momentum different but still close to zero, strongly increase their energy. Although the replacement of composite bosons by elementary bosons noticebly affects these exchanges, we have recently shown \cite{CombescotSnoke} that the same argument for the non-fragmentation of Bose-Einstein condensate also holds for composite bosons.

Although counterintuitive, it is now clear that after many years of search, Bose-Einstein condensation (BEC) of boson-like particles is not a textbook effect anymore: it is commonly observed for a large variety of atoms\cite{Anderson,Davis,Bradley,Fried,Robert,Cornish,Weber} and molecules\cite{Wynar,Jochim}, as seen from the exploding field of "cold quantum gases".
While semiconductor excitons have early appeared as good candidates to observe this condensation and a large amount of experimental work has been devoted to this observation\cite{Nagasawa,Nakata,Chase,Hasuo1993,Hasuo1994,Lin,Mysyrowicz,Shen,Fukuzawa,Pau,Butov1994,Butov2002,SnokeNature2002,Larionov,Snoke2002}, all claims for exciton BEC have been inconclusive\cite{Snoke2003}. One of the most puzzling phenomena so far is the observation of ring-like structures in the photoluminescence of indirect excitons containing a dark region\cite{Butov2002,SnokeNature2002}.

Very recently, experimentalists have turned to polaritons and demonstrated their BEC in semiconductor quantum well embedded inside a microcavity\cite{Kasprzak,Deng2002,Deng2003,Deng2006}. It was however argued that the observed coherence could be a bare lasing effect\cite{Snoke2006}. In more recent experiments, a trap is used by means of a rounded-tip pin and polaritons are produced by an incoherent source. They then move to the trap where they condensate, thus providing clearer evidence of polariton condensation\cite{Balili}.

Polaritons are quite complex composite bosons. This makes their interactions uneasy to handle \cite{CombescotEPL}. Being eigenstates of one photon strongly coupled to one exciton, polaritons are a linear combination of photon and exciton with prefactors depending on the polariton momentum. They thus partly are elementary boson -- through their photon component -- and composite boson -- through their exciton component. It can then be argued that the photon part of these polaritons could greatly facilitate the condensation, $k=0$ polaritons having a quite large photon part which makes them very light.

In contrast, excitons are composite bosons; far more similar to atoms than polaritons. Two puzzling questions thus arise: Why has the exciton condensation not been observed yet? What is the dark region in the above-mentioned photoluminescence spectra \cite{Butov2002,SnokeNature2002}? Strangely enough, only last year we have most probably caught the point \cite{CombescotCombescot}. Excitons are created by photon absorption in states which are by construction bright. However, due to their composite nature, excitons not only interact by Coulomb processes but also by carrier exchanges. It becomes immediately clear that a pair of bright excitons with opposite spins (+1,-1) can be transformed into a pair of dark excitons (+2,-2) by means of carrier exchange (see Fig.~\ref{2Xexchange}), possibly mixed with Coulomb interaction.

\begin{figure}[htp]
\onefigure[width=8cm]{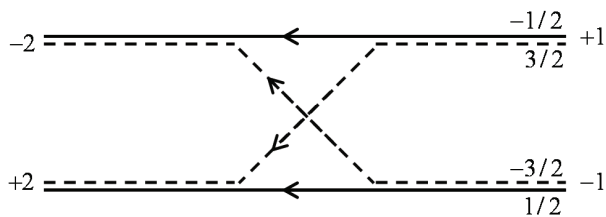}
\caption{Carrier exchange between two excitons which converts bright excitons into dark excitons. Solid lines: electrons, dashed lines: holes.}
\label{2Xexchange}
\end{figure}

Consequently, even if excitons are created in bright states, a similar amount of dark excitons are formed on a characteristic time determined by the inverse of the Coulomb exchange energy $R_xN(a_x/L)^D$, where $N$ is the exciton number, $R_x$ and $a_x$ are the exciton Rydberg and Bohr radius, $L$ is the sample size, and $D$ the space dimension. Note that this time is also the characteristic time for reaching equilibrium among bright excitons, since the direct scattering of two excitons with close to 0 momenta is essentially zero \cite{CombescotPR}. 
While bright and dark excitons are degenerate in energy if we only take into account Coulomb {\it intraband direct} processes, dark excitons become the energetically lowest states if Coulomb {\it interband exchange} processes are included, since spin conservation in the valence-conduction transition makes these processes possible for $(0,\pm 1)$ excitons, but not for $(\pm 2)$ excitons. As Coulomb interband exchange interaction between valence and conduction electrons is repulsive, dark excitons end up having a slightly smaller energy than bright excitons. Note that Coulomb intraband direct interaction between two electrons also is repulsive, but the one between a conduction electron and a valence electron absence, i.e. a hole, is attractive, this attraction leading to the binding of excitons.
Indeed, for $q\ne 0$, {\it interband} transition is governed by
\bea
a\D_{v,\bk+\bq}a\D_{c,\bk'-\bq}a_{v,\bk'}a_{c,\bk} & = & b_{-\bk-\bq}a\D_{\bk'-\bq}b\D_{-\bk'}a_\bk \nn\\
& = & a\D_{\bk'-\bq}b\D_{-\bk'}b_{-\bk-\bq}a_\bk,
\eea
while in the case of {\it intraband} transition, we do have
\bea
a\D_{c,\bk+\bq}a\D_{v,\bk'-\bq}a_{v,\bk'}a_{c,\bk} & = & a\D_{\bk+\bq}b_{-\bk'-\bq}b\D_{-\bk'}a_\bk \nn\\
& = & -a\D_{\bk+\bq}b\D_{-\bk'}b_{-\bk'-\bq}a_\bk,
\eea
the conduction/valence and electron/hole operators being related by $a\D_{c,\bk}=a\D_\bk$ and $a\D_{v,\bk}=b_{-\bk}$. 

We thus conclude that, if excitons have to condense, they must condense in dark states. Up to now, all experiments aiming at showing evidence for the condensation of excitons have made use of luminescence spectroscopy. This actually is rather strange because the fact that dark excitons are lower in energy than bright excitons was widely known \cite{Pikus,Chen,Kesteren,Maialle,Blackwood,Bayer}. A possible reason for missing the point could be that the Shiva diagram for the carrier exchange shown in Fig.~\ref{2Xexchange} never appeared in the literature up to our works on composite exciton many-body effects \cite{CombescotSSC,CombescotPR}, and therefore people may not have realized that these low energy dark excitons are present in their system, through whatever method they create excitons. It seems plausible that in the experiments performed in Refs.~\cite{Butov2002,SnokeNature2002}, the observed dark region inside the luminescent ring is in reality made of $(\pm 2)$ excitons. 

This very straightforward remark about carrier exchange between bright states sheds new light on the whole field of exciton Bose-Einstein condensation. A major problem still remains: If the exciton condensate is dark, how can we "see" it? An idea is to use a trap because if the condensate has to be formed, it is going for sure to condense in the lowest level of the trap. Since trap levels usually have different spatial extensions, we thus expect the center of the trap to  appear dark when condensation takes place.

\begin{figure}[htp]
     \centering
     \subfigure[Photon absorption]{
          \label{Vphoton_a}
          \includegraphics[width=35mm]{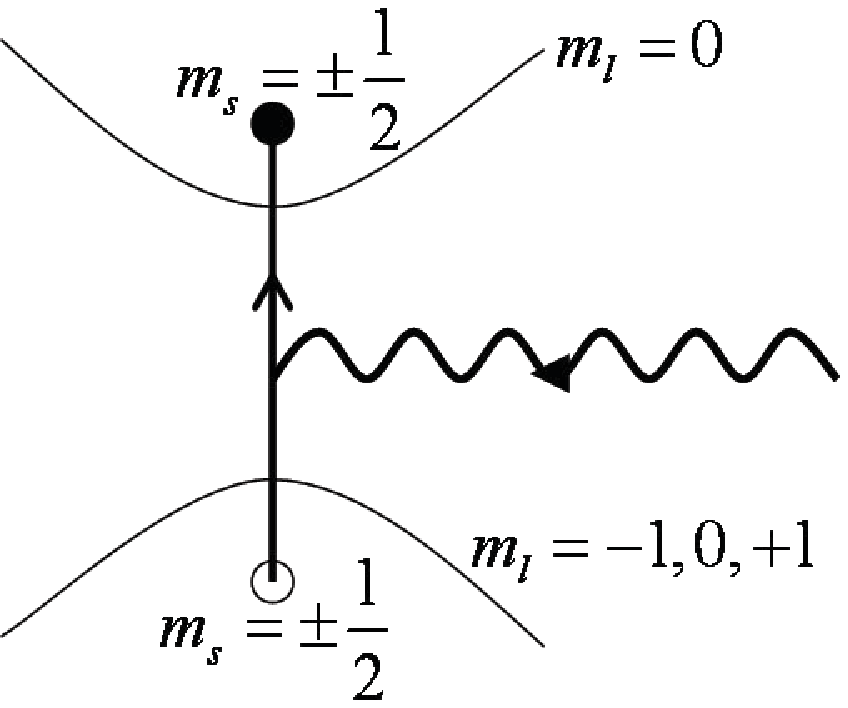}}
     \hspace{4mm}
     \subfigure[Photon emission]{
          \label{Vphoton_b}
          \includegraphics[width=35mm]{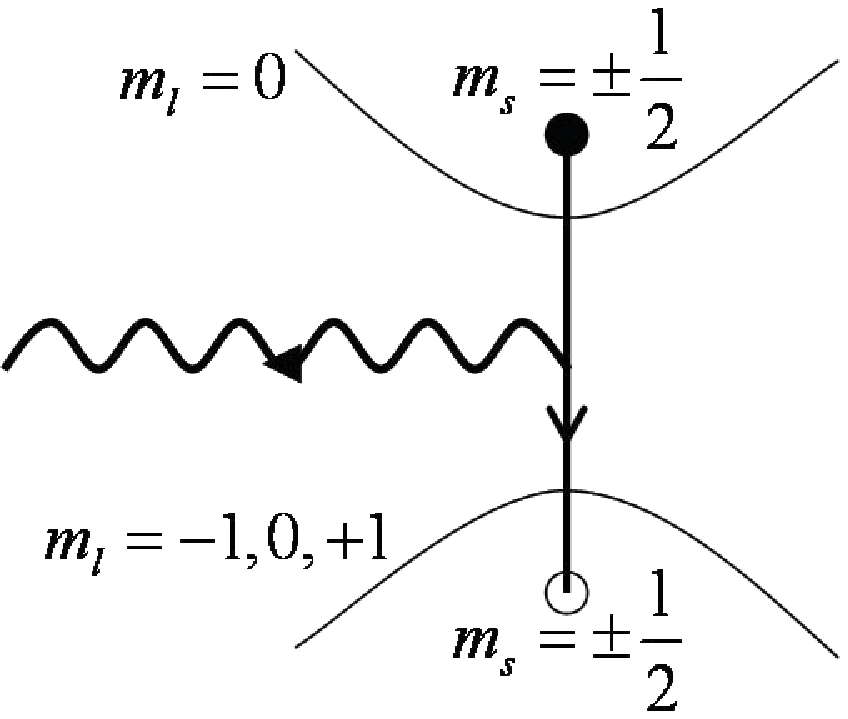}}\\
     \subfigure[Photon absorption]{
           \label{Vphoton_c}
          \includegraphics[width=42mm]{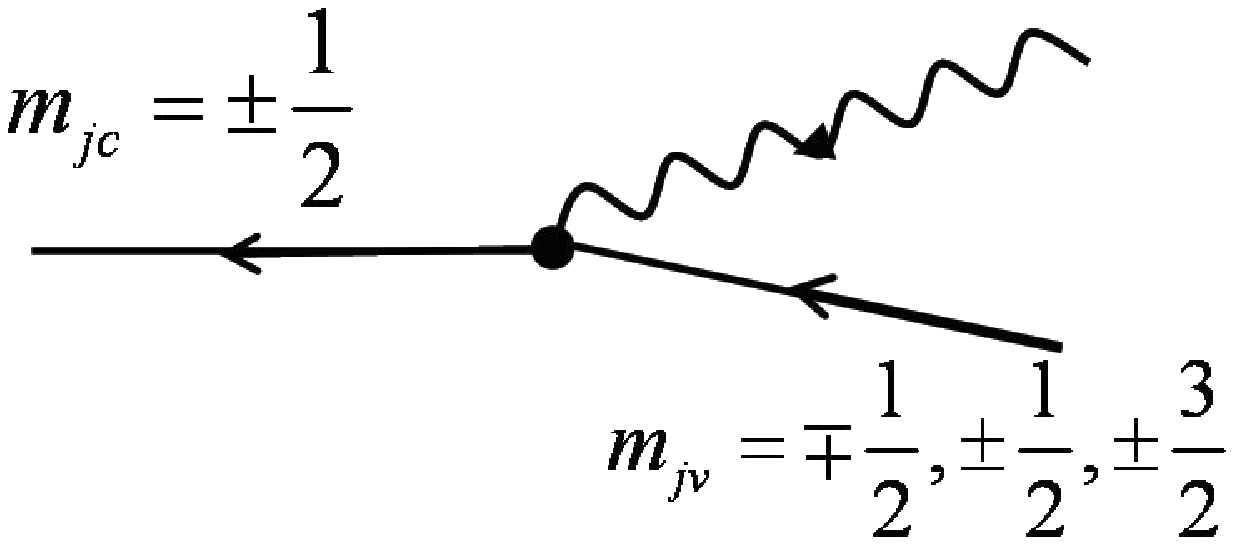}}
     \subfigure[Photon emission]{
          \label{Vphoton_d}
          \includegraphics[width=42mm]{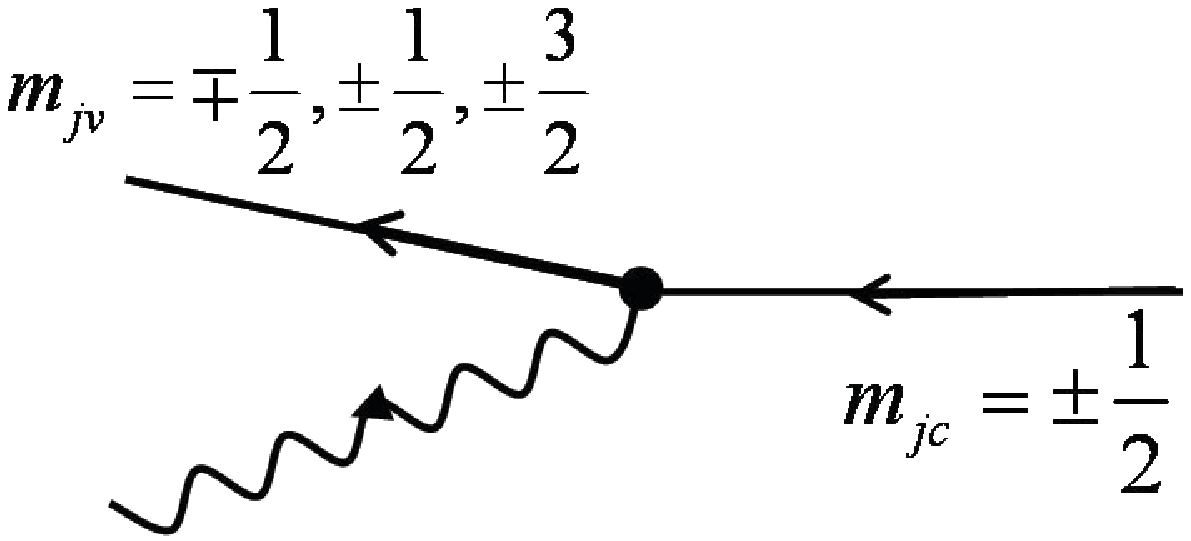}}
     \caption{Photon absorption (a) and emission (b): Spin is conserved in the valence-conduction transition. In (c), one photon plus one valence electron gives one conduction electron, while in (d), one conduction electron gives one photon plus one valence electron, the electron angular momenta being $m_j=m_s+m_l$.}
     \label{Vphoton}
\end{figure}

We can first think of a trap in a single quantum well. However, in order to have a long exciton lifetime, as necessary to reach a density high enough for BEC at temperatures possibly obtained in photoexcited exciton gas, a double quantum well, with electrons and holes spatially separated, seems more appropriate than a single quantum well.
Since the standard interband Coulomb exchange matrix element, which makes bright states lie energetically above dark states, contains the product of electron and hole wavefunctions, it becomes crucial to base our argument on more general grounds, in order to possibly claim that, even in the absence of apparent electron-hole spatial overlap, excitons should condense in dark states not only for a single but also for a double quantum well, in which electrons and holes are separated from each other.

We now develop this general argument. It shows that for the very same reason that excitons are bright, their energy is higher than the dark exciton energy. For that, let us first come back to the reason for which $(\pm 1,0)$ excitons are bright, the link with interband Coulomb processes then becoming transparent. 

The angular momentum selection rule for the absorption and emission of one photon leading to the valence-conduction transitions  shown in Figs.~\ref{Vphoton_a} and \ref{Vphoton_b} are due to spin conservation. In the case of absorption, the electron changes its orbital angular momentum from $(l=1,m_l=-1,0,1)$ to $(l=0,m_l=0)$ while keeping its spin $m_s=\pm 1/2$; so that, in terms of valence-conduction electrons, this transition is associated with the diagram of Fig.~\ref{Vphoton_c}, the total electron angular momentum projected on the axis perpendicular to the quantum well being $m_j=m_l+m_s$; and similarly for photon emission shown in Fig.~\ref{Vphoton_d}. We can now change from conduction-valence electron to electron-hole description, as shown in Figs.~\ref{Photoneh_a} and \ref{Photoneh_b}, the hole angular momentum being $m_{jh}=-m_{jv}$. This readily shows that the electron-hole pair generated by one photon has a total angular momentum $M_{jeh}=m_{je}+m_{jh}=(\pm 1,0)$, which correspond to transitions possibly induced by $(\sigma_\pm,\pi)$ photons, respectively.
In the same way, the conduction-valence transition shown in Fig.~\ref{Vphoton_d} corresponds to the annihilation of one electron-hole pair (see Fig.~\ref{Photoneh_b}), again having a total angular momentum $M_{jeh}=m_{je}+m_{jh}=(\pm 1,0)$, and thus also coupled to $(\sigma_\pm,\pi)$ photons. This allows us to recover the well-known fact that creation or annihilation of one electron-hole pair $(\pm 1,0)$ in semiconductors is associated to the absorption or emission of one photon. 

\begin{figure}[htp]
     \centering
     \subfigure[Photon absorption]{
          \label{Photoneh_a}
          \includegraphics[width=42mm]{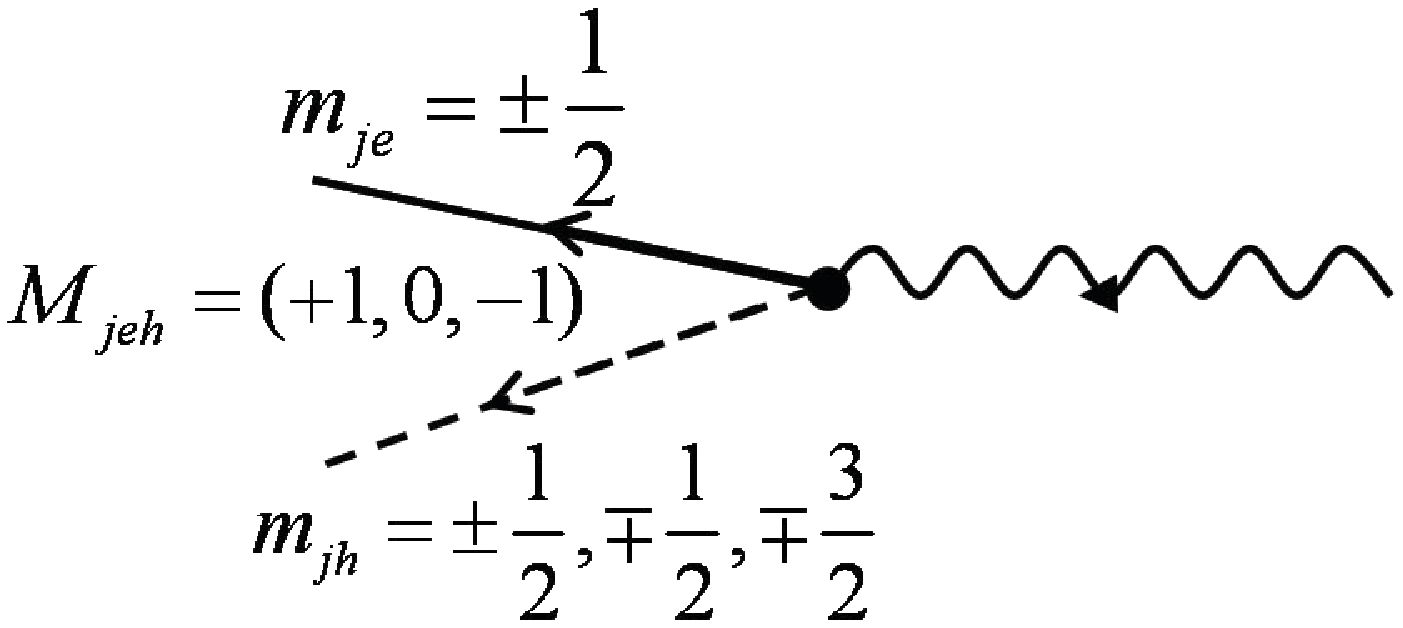}} 
     \subfigure[Photon emission]{
           \label{Photoneh_b}
           \includegraphics[width=42mm]
                {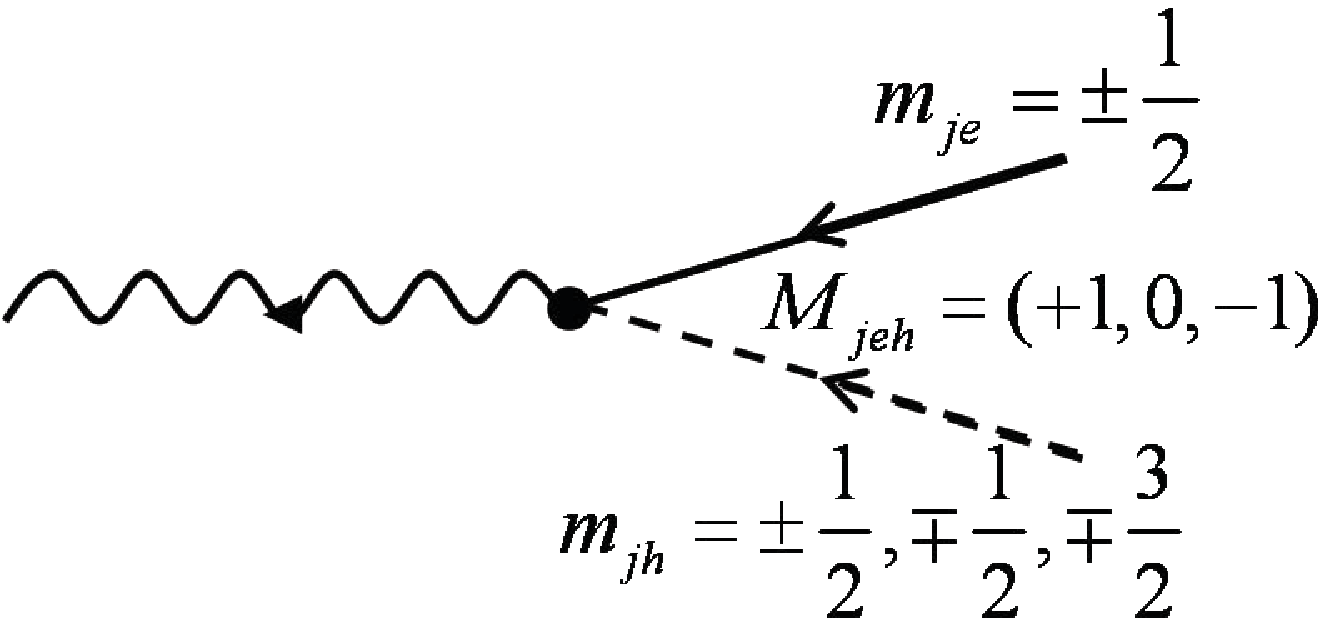}}
     \caption{Same as diagrams of Figs.~\ref{Vphoton_c} and \ref{Vphoton_d}, respectively, in terms of electron and hole, with $m_{je}=m_{jc}$, while $m_{jh}=-m_{jv}$. We see that the electron-hole pairs which are created or recombine, i.e. the pairs which are bright, have a total angular momentum $(\pm 1,0)$.}
     \label{Photoneh}
\end{figure}

Let us now consider the Coulomb {\it interband exchange} interaction shown in Figs.~\ref{Interbandexchange_a} or \ref{Interbandexchange_b}, in which one conduction electron in state $c_1$ goes to the valence band in state $v_1$, while one valence electron in state $v_2$ goes to the conduction band in state $c_2$. By rewriting the diagram \ref{Interbandexchange_c} in terms of electron and hole as in \ref{Interbandexchange_d}, it becomes obvious that this valence-conduction Coulomb exchange process is just the succession of emission and absorption of one virtual photon, a fact well known in relativistic quantum field theory\cite{Sakurai}.
The relativistic counterpart of the diagram shown in Fig.~\ref{Interbandexchange_b} is called the Bhabha exchange scattering in the valence-conduction electron picture, which corresponds to the M{\o}ller exchange scattering between two electrons. In the electron-hole picture, this Coulomb exchange interaction corresponds to the Bhabha annihilation scattering \cite{Sakurai}, which has the opposite sign of the direct scattering and is therefore repulsive between electrons and holes. 
From the above argument, we thus conclude that interband Coulomb exchanges only exist for excitons made of $(\pm 1,0)$ electron-hole pairs, those excitons coupled to photons being thus bright.

Coulomb interaction also contains {\it intraband direct} processes (Fig.~\ref{Intraband_a}) between one conduction and one valence electron (Fig.~\ref{Intraband_b}), or equivalently between one electron and one hole (Fig.~\ref{Intraband_c}). When repeated, we get the set of ladder processes for electron-hole pair shown in Fig.~\ref{Intraband_d}, which is responsible for the formation of excitons\cite{Fetter}. The corresponding sum leads to the exciton energy spectrum which, for these intraband Coulomb processes, does not depend on electron and hole spin. At this level, bright excitons $(\pm 1,0)$ and dark excitons $(\pm 2)$ would be degenerate.

\begin{figure}[htp]
     \centering
     \subfigure[Interband Coulomb exchange]{
          \label{Interbandexchange_a}
          \includegraphics[width=35mm]{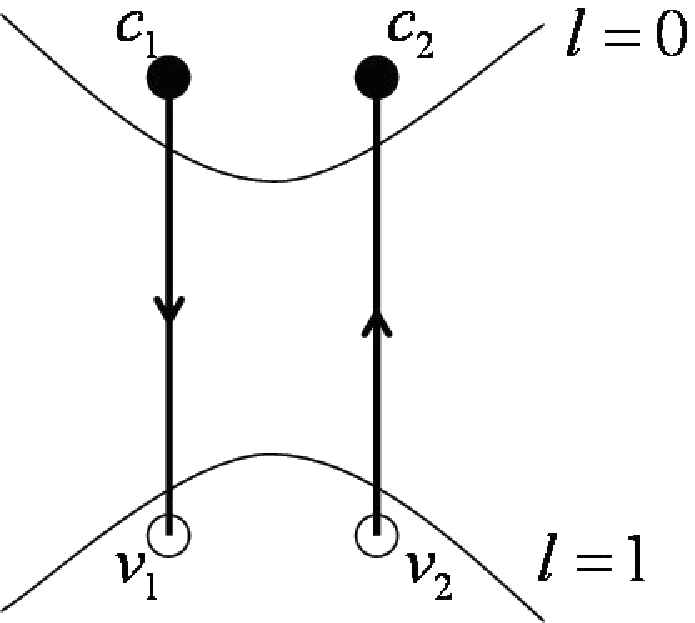}}
     \hspace{4mm}
     \subfigure[Valence-conduction exchange diagram]{
           \label{Interbandexchange_b}
           \includegraphics[width=38mm]{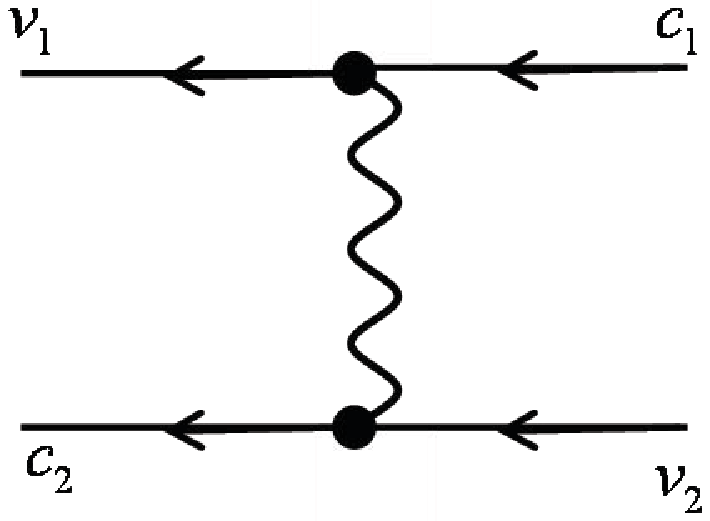}}\\
     \subfigure[Rearranging the exchange diagram (b)]{
          \label{Interbandexchange_c}
          \includegraphics[width=60mm]{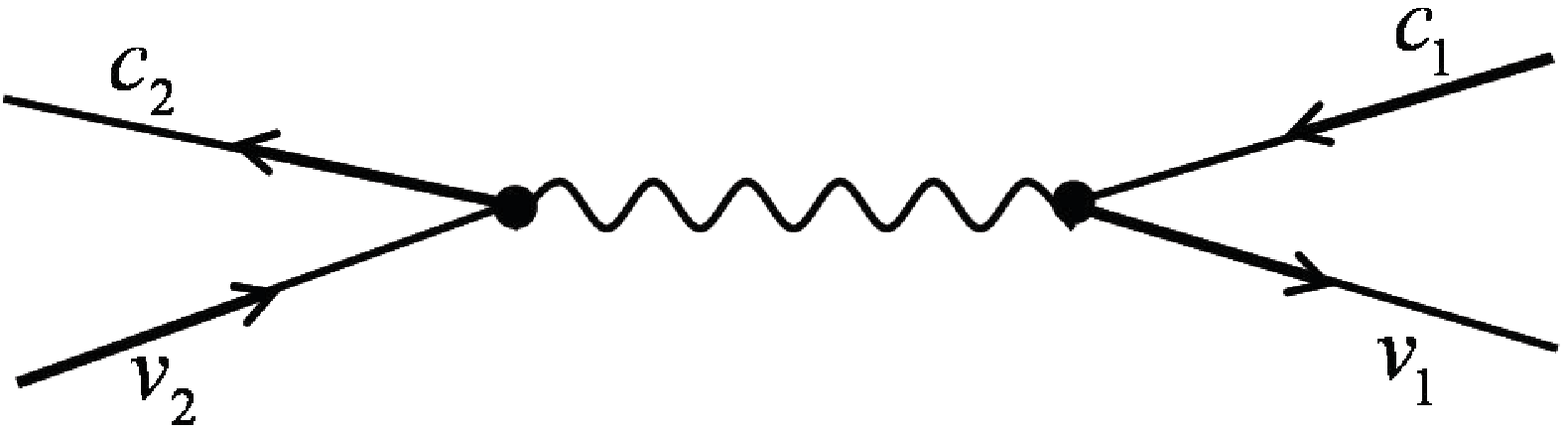}}\\
     \subfigure[Exchange diagram with electron-hole]{
           \label{Interbandexchange_d}
           \includegraphics[width=8cm]
                {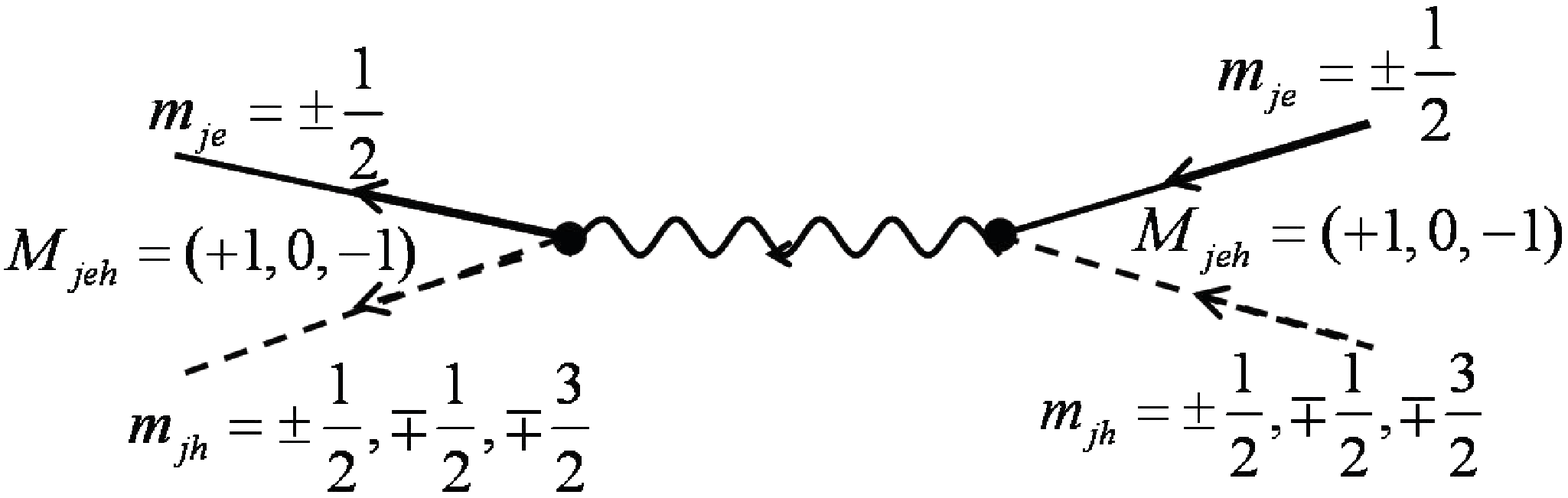}}
     \caption{Interband Coulomb exchange interaction which shifts the bright exciton energy above the dark exciton energy. Transition process either in terms of valence-conduction electrons (b) and (c), or in terms of electron-hole (d). When compared to Fig.~\ref{Vphoton}, we readily see that this interband Coulomb process is nothing but an exchange of virtual photon between electron-hole pairs.}
     \label{Interbandexchange}
\end{figure}

If we now remember that bright excitons $(\pm 1,0)$ can also undergo Coulomb interband exchange processes, we must for these excitons add the diagram of Fig.~\ref{Xexchange}. 
As the interband Coulomb exchange interaction is repulsive and obeys the above-mentioned selection rules for the production of a virtual photon, we readily conclude that additional contributions like the one of Fig.~\ref{Xexchange} 
push bright excitons up in energy as opposed to dark exciton states, which do not have these interband processes. Since this argument makes only use of the fact that in valence-conduction transition, the electron keeps its spin, it is thus valid for any spatial configuration, i.e. for single or double quantum well as well as for sample under stress, such as in the case of a trap made by a pin. 

It can be argued that the matrix element for interband Coulomb process generated by the standard second quantization procedure reads in terms of integral in which the wave functions of "in" and "out" electron states appear as a product. In a double well configuration, we could naively conclude that this product reduces to zero, electrons being in one well and holes in the other well. However, while a similar second quantization procedure makes the same wavefunction product enter the matrix element appearing in the semiconductor-photon interaction, we do see bright excitons in a double quantum well. This proves that in a double quantum well an indirect channel must exist which leads to a non-zero {\it effective} overlap between electron and hole wavefunctions, as necessary to observe exciton photoluminescence through recombination of electron-hole pairs. {\it The same channel} then produces the non-zero Coulomb interband exchange interaction which pushes bright exciton above dark exciton states. It is however clear that this effective overlap is going to be much smaller for a double quantum well than for the direct overlap which exists for a single quantum well. This reduction, which actually is the reason for having an increased exciton lifetime  compared to a single quantum well, makes the bright-dark exciton splitting reduced for a double quantum well. This reduction, obviously dependent on the well at hand, is clearly quite difficult to calculate precisely. Fortunately, its precise value is unimportant for dark exciton BEC since it is not the size of the energy splitting which produces Bose-Einstein condensation, a fact evidenced by the condensation of $k=0$ bosonic atoms, extremely close in energy to those with momentum $2\pi/L$ for large sample size $L$.

\begin{figure}[htp]
     \centering
     \subfigure[Coulomb intraband interaction]{
          \label{Intraband_a}
          \includegraphics[width=35mm]{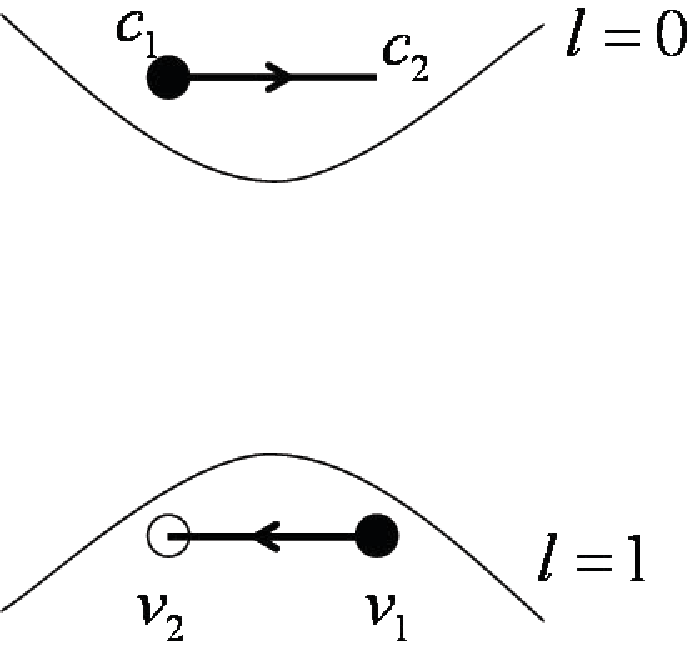}}
     \hspace{4mm}
     \subfigure[direct interaction with valence-conduction electrons]{
          \label{Intraband_b}
          \includegraphics[width=35mm]{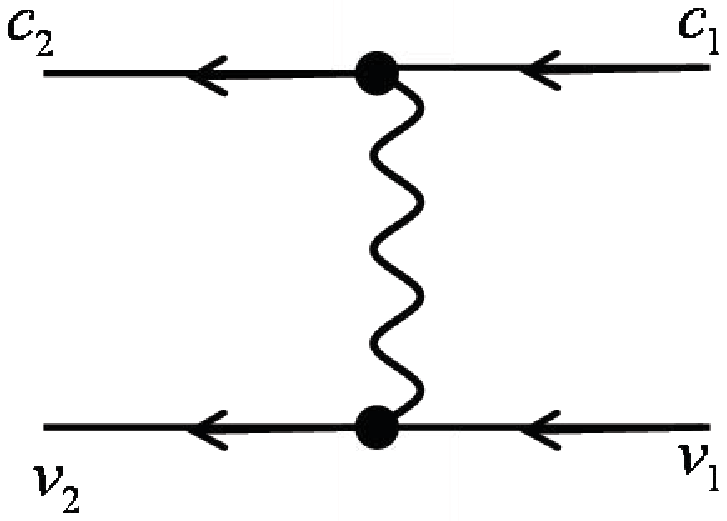}} \\
     \vspace{.3in}
     \subfigure[direct interaction with electron-hole]{
           \label{Intraband_c}
           \includegraphics[width=35mm]
                {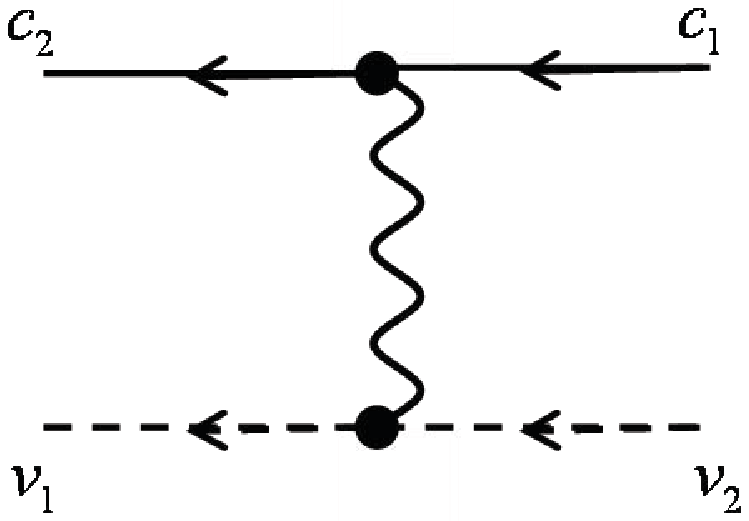}}
	\hspace{4mm}
     \subfigure[exciton ladder processes]{
          \label{Intraband_d}
          \includegraphics[width=35mm]{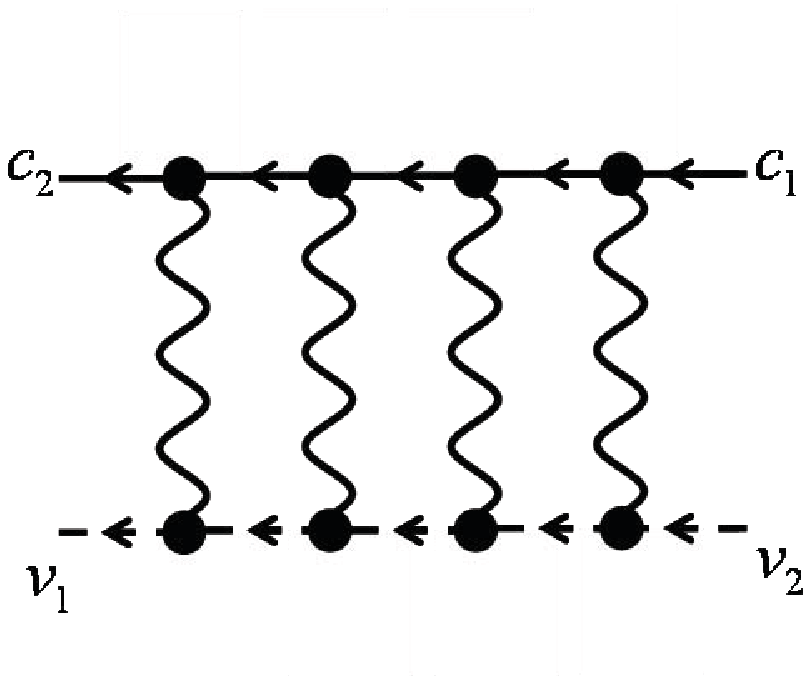}}
     \caption{(a) Coulomb intraband interaction: (b) one conduction electron goes from $c_1$ to $c_2$, while one valence electron goes from $v_1$ to $v_2$, or equivalently, (c) one hole goes the other way from $v_2$ to $v_1$. (d) Set of ladder diagrams between one electron and one hole, leading to exciton.}
     \label{Intraband}
\end{figure}

The above discussion leads us to conclude that just because $(0,\pm 1)$ excitons are bright, i.e. coupled to photons, they have a higher energy than $(\pm 2)$ dark excitons, whatever the electron-hole spatial configuration is. Consequently, even if the bright/dark exciton splitting is decreased in a double quantum well structure -- for the same reason that the bright exciton lifetime is increased -- Bose-Einstein condensates of excitons have to be made out of dark states in double quantum wells, too. Exciton BEC can then be evidenced as the appearance of a dark spot at the center of a trap, when single or double quantum well samples are cooled down.

\begin{figure}[htp]
\onefigure[width=8cm]{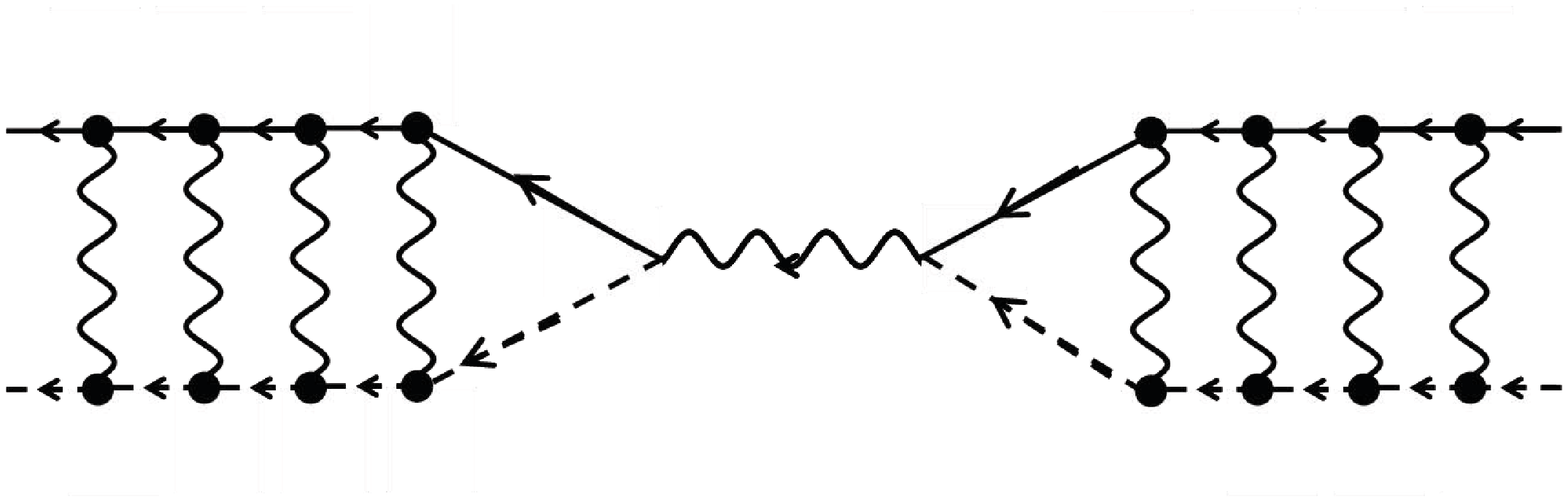}
\caption{Interband Coulomb exchange interaction between electron and hole pushing bright excitons $(\pm 1,0)$ above dark excitons $(\pm 2)$.}
\label{Xexchange}
\end{figure}

Let us note that in the Coulomb exchange scattering which transforms two bright excitons $(+1,-1)$ into two dark excitons $(+2,-2)$, energy must be conserved. Since bright excitons are above in energy, this means that, for bright excitons with close to zero momentum, as obtained by photon absorption, the resulting dark excitons have momenta $(+Q^*,-Q^*)$ with ${Q^*}^2/2M_X=E_{+1}-E_{+2}$. These dark excitons can then lose their magic momentum $Q^*$ through similar Coulomb exchange scatterings but with same spin dark excitons, so that they stay dark.
Through these scatterings, some of them end by having $Q=0$ momentum to possibly condense. 

With respect to the optically observed polariton BEC we can say that, since photons get trapped inside the cavity and interact strongly with bright excitons, the new particle made of a superposition of photon and bright exciton, i.e. the polariton, has a lower branch in the dispersion that lies below the dark exciton state. At $k=0$ the photonic component of the polariton is very strong, which makes the polariton very light. BEC then takes place for $k=0$ polaritons in the lower branch.

We can also try to understand the ring structure observed in photoluminescence experiments \cite{Butov2002,SnokeNature2002} within the same dark exciton framework. One key can be that carrier exchange transforms pairs of bright excitons with spins (+1,-1) into pairs of dark excitons with spins (+2,-2). In GaAs quantum wells, the bright-dark splitting is of the order of 0.1 meV \cite{Blackwood}, which corresponds to about 1 K. For double quantum well with electrons and holes spatially separated, we expect a far shorter exchange splitting, of the order of $1$ $\mu$eV or less \cite{Kesteren}. Such a small splitting is not a problem for having dark exciton BEC because condensation can take place even for a very small splitting, as evidenced by BEC of $k=0$ atoms or $k=0$ polaritons in macroscopic samples.


When the temperature is low enough and the density high enough, $k=0$ dark excitons are in fact locked in a dark condensate, being prevented from transforming back into bright excitons due to quantum effects, i.e. lack of available state population at the temperature of the experiment. In the absence of trap, these dark excitons can however move radially away from the bright pump spot where they are created by carrier exchange, their density gradually decreasing. At a critical density, the one presumably reached at the outer ring radius, the dark exciton condensate must evaporate, unlocking excitons to be transformed back into bright states. This could be an explanation for the observed luminescent outer ring. Since the motion of the dark BEC must be coherent, the fluctuations of the radius of the outer ring has to be very small. This density argument used for the appearence of the ring is also consistent with the experimental observation of the merging of two rings in the case of two pump spots \cite{Snoke2003}. The appearance of luminescent spots inside the ring clearly needs some additional effect. While a regular distribution of spots would push toward a diffraction-like pattern, less regular positions could be interpreted as being due to inhomogeneities of the quantum well width acting as tiny traps: a single monolayer change induces an energy difference of the order of 1 meV, which is large compared to the dark-bright exciton energy splitting. If the lateral size of the confinement in these traps is much smaller than the exciton Bohr radius $a_x$, the Coulomb interaction is negligible compared to confinement, thus leading to degeneracy between bright and dark excitons. In addition, these trapped excitons are decoupled from the condensate. Therefore, they can recombine, which could explain the observed bright spots.






In conclusion, by just using the fact that electrons keep their spin in Coulomb process, we recover the fact that interband Coulomb exchange interaction is nothing but a sequence of emission and absorption of one virtual photon. This repulsive Coulomb process thus pushes excitons coupled to photons, i.e. bright excitons, above dark excitons, whatever the electron-hole spatial configuration is. Consequently, excitons should condense in dark states for a single as well as a double quantum well, i.e. for a structure having a long exciton lifetime: This dark condensate can actually form for arbitrary but finite dark-bright exciton splitting. If a trap is to be made in these wells, we thus predict exciton BEC to appear as a dark spot at the trap center. The bright ring observed in photoluminescence experiments could also be explained along the same idea, although the observed bright spots inside the ring need some mixed effects.

\acknowledgments
One of us (M.C.) wishes to thank David Snoke for pointing out the interest in double quantum well structures. The other (M.N.L.) acknowledges partial support through the NSF under Grant No. ECCS 0725514 and through the DARPA/MTO Young Faculty Award under Grant No. HR0011-08-1-0059.


\begin{thebibliography}{0}

\bibitem{Pitaevskii}
  \Name{Pitaevskii L. \and Stringari S.}
  \Book{Bose-Einstein Condensation}
  \Publ{Oxford University Press}
  \Year{2003}

\bibitem{Nozieres}
  \Name{Nozieres P.}
  \Book{Bose-Einstein Condensation}
  \Editor{Griffin A., Snoke D. W. \and Stringari S.}
  \Publ{Cambridge University Press}
  \Year{1995}

\bibitem{Leggett}
  \Name{Leggett A. J.}
  \Book{Quantum Liquids: Bose-Einstein Condensation and Cooper pairing in condensed matter systems}
  \Publ{Oxford University Press}
  \Year{2008}

\bibitem{CombescotSnoke}
  \Name{Combescot M. \and Snoke D.}
  \REVIEW{submitted to Phys. Rev. B}{}{}{}.

\bibitem{Anderson}
  \Name{Anderson M. H., Ensher J. R., Matthews M. R., Wieman C. E. \and Cornell E. A.}
  \REVIEW{Science}{269}{1995}{198}.

\bibitem{Davis}
  \Name{Davis K. B. \etal}
  \REVIEW{Phys. Rev. Lett.}{75}{1995}{3969}.

\bibitem{Bradley}
  \Name{Bradley C. C., Sackett C. A., Tollet J. J. \and Hulet R. G.}
  \REVIEW{Phys. Rev. Lett.}{75}{1995}{1687}.

\bibitem{Fried}
  \Name{Fried D. G. \etal}
  \REVIEW{Phys. Rev. Lett.}{81}{1998}{3811}.

\bibitem{Robert}
  \Name{Robert A. \etal}
  \REVIEW{Science}{292}{2001}{461}.

\bibitem{Cornish}
  \Name{Cornish S. L. \etal}
  \REVIEW{Phys. Rev. Lett.}{85}{2000}{1795}.

\bibitem{Weber}
  \Name{Weber T., Herbig J., Mark M., N\"agerl H.-C. \and Grimm R.}
  \REVIEW{Science}{299}{2003}{232}.

\bibitem{Wynar}
  \Name{Wynar R., Freeland R. S., Han D. J., Ryu C. \and Heinzen D. J.}
  \REVIEW{Science}{287}{2000}{1016}.

\bibitem{Jochim}
  \Name{Jochim S. \etal}
  \REVIEW{Science}{301}{2003}{2101}.

\bibitem{Nagasawa}
  \Name{Nagasawa N., Nakata N., Doi Y. \and Ueta J.}
  \REVIEW{J. Phys. Soc. Jpn.}{38}{1975}{593}.

\bibitem{Nakata}
  \Name{Nakata N., Nagasawa N., Doi Y. \and Ueta J.}
  \REVIEW{J. Phys. Soc. Jpn.}{38}{1975}{903}.

\bibitem{Chase}
  \Name{Chase L. L., Peyghambarian N., Grinberg G. \and Mysyrowicz A.}
  \REVIEW{Phys. Rev. Lett.}{42}{1979}{1231}.

\bibitem{Hasuo1993}
  \Name{Hasuo M., Nagasawa N., Itoh T. \and Mysyrowicz A.}
  \REVIEW{Phys. Rev. Lett.}{70}{1993}{1303}.

\bibitem{Hasuo1994}
  \Name{Hasuo M., Nagasawa N., Itoh T. \and Mysyrowicz A.}
  \REVIEW{J. Lumin.}{60/61}{1994}{758}.

\bibitem{Lin}
  \Name{Lin J. L. \and Wolfe J. P.}
  \REVIEW{Phys. Rev. Lett.}{71}{1993}{1222}.

\bibitem{Mysyrowicz}
  \Name{Mysyrowicz A., Benson E. \and Fortin E.}
  \REVIEW{Phys. Rev. Lett.}{77}{1996}{896}.

\bibitem{Shen}
  \Name{Shen M. Y., Yokouchi T., Koyama S. \and Goto T.}
  \REVIEW{Phys. Rev. B}{56}{1997}{13066}.

\bibitem{Fukuzawa}
  \Name{Fukuzawa T., Mendez E. E. \and Hong J. M.}
  \REVIEW{Phys. Rev. Lett.}{64}{1990}{3066}.

\bibitem{Pau}
  \Name{Pau S., Cao H., Jacobson J., Bj\"ork G. \and Yamamoto Y.}
  \REVIEW{Phys. Rev. A}{54}{1996}{R1789}.

\bibitem{Butov1994}
  \Name{Butov L. V., Zrenner A., Abstreiter G., B\"ohm G. \and Weimann G.}
  \REVIEW{Phys. Rev. Lett.}{73}{1994}{304}.

\bibitem{Butov2002}
  \Name{Butov L. V., Gossard A. C. \and Chemla D. S.}
  \REVIEW{Nature}{418}{2002}{751}.

\bibitem{SnokeNature2002}
  \Name{Snoke D., Denev S., Liu Y., Pfeiffer L. \and West K.}
  \REVIEW{Nature}{418}{2002}{754}.

\bibitem{CombescotCombescot}
  \Name{Combescot M., Betbeder-Matibet O. \and Combescot R.}
  \REVIEW{Phys. Rev. Lett.}{9}{2007}{176403}.

\bibitem{Larionov}
  \Name{Larionov A. V., Timofeev V. B., Hvam J. \and Soerensen K.}
  \REVIEW{JETP}{90}{2000}{1093}.

\bibitem{Snoke2002}
  \Name{Snoke D.}
  \REVIEW{Science}{298}{2002}{1368}.

\bibitem{Snoke2003}
  \Name{Snoke D.}
  \REVIEW{Phys. Stat. Sol. (b)}{238}{2003}{389}.

\bibitem{Kasprzak}
  \Name{Kasprzak \etal}
  \REVIEW{Nature}{443}{2006}{409}.

\bibitem{Deng2002}
  \Name{Deng H., Weihs G., Santori C., Bloch J. \and Yamamoto Y.}
  \REVIEW{Science}{298}{2002}{199}.

\bibitem{Deng2003}
  \Name{Deng H., Weihs G., Snoke D. W., Bloch J. \and Yamamoto Y.}
  \REVIEW{Proc. Natl. Acad. Sci. U.S.A.}{100}{2003}{15318}.

\bibitem{Deng2006}
  \Name{Deng H. \etal}
  \REVIEW{Phys. Rev. Lett.}{97}{2006}{146402}.

\bibitem{Snoke2006}
  \Name{Snoke D.}
  \REVIEW{Nature}{443}{2006}{403}.

\bibitem{Balili}
  \Name{Balili R., Hartwell V., Snoke D., Pfeiffer L. \and West K.}
  \REVIEW{Science}{316}{2007}{1007}.

\bibitem{CombescotEPL}
  \Name{Combescot M., Dupertuis M. A. \and Betbeder-Matibet O.}
  \REVIEW{Europhys. Lett.}{79}{2007}{17001}.

\bibitem{Fetter}
  \Name{Fetter A. L. \and Walecka J. D.}
  \Book{Quantum Theory of Many-Particle Systems}
  \Publ{McGraw-Hill, New York}
  \Year{1971}

\bibitem{Pikus}
  \Name{Pikus G. E. \and Bir G. L.}
  \REVIEW{Soviet Phys. JETP}{33}{1971}{108}.

\bibitem{Chen}
  \Name{Chen Y., Gil B., Lefebvre P. \and Mathieu H.}
  \REVIEW{Phys. Rev. B}{37}{1988}{6429}.

\bibitem{Kesteren}
  \Name{van Kesteren H. W., Cosman E. C., van der Poel W. A. J. A. \and Foxon C. T.}
  \REVIEW{Phys. Rev. B}{41}{1990}{5283}.

\bibitem{Maialle}
  \Name{Maialle M. Z., de Andrada e Silva E. A. \and Sham L. J.}
  \REVIEW{Phys. Rev. B}{47}{1993}{15776}.

\bibitem{Blackwood}
  \Name{Blackwood E., Snellilng M. J. \and Harley R. T.}
  \REVIEW{Phys. Rev. B}{50}{1994}{14246}.

\bibitem{Bayer}
  \Name{Bayer M. \etal}
  \REVIEW{Phys. Rev. B}{65}{2002}{195315}.

\bibitem{CombescotSSC}
  \Name{Combescot M. \and Betbeder-Matibet O.}
  \REVIEW{Sol. State Comm.}{134}{2005}{11}.

\bibitem{CombescotPR}
  \Name{Combescot M.,  Betbeder-Matibet O. \and Dubin F.}
  \REVIEW{Phys. Rep.}{463}{2008}{215}.

\bibitem{Sakurai}
  \Name{Sakurai J. J.}
  \Book{Advanced Quantum Mechanics}
  \Publ{Addison-Wesley, Massachussetts}
  \Year{1967}


\end{thebibliography}
\end{document}